\documentclass[prb, reprint, aps, superscriptaddress, showpacs, floatfix]{revtex4-2}
\usepackage[margin=30truemm]{geometry}
\usepackage [dvipdfmx]{graphicx}
\usepackage{float}						
\usepackage{wrapfig}						
\usepackage{ascmac}
\usepackage{amsmath}
\usepackage{bm}
\usepackage{comment}
\usepackage{cases}
\usepackage{amssymb}
\usepackage{mathtools}
\usepackage{fancybox}
\usepackage{CJK}
\usepackage{color}
\usepackage{pstricks}
\usepackage{cancel}						
\usepackage{braket}						
\usepackage{amsthm} 
\usepackage{lipsum}
\usepackage{indentfirst}

\begin{document}
\title{Realization of topological phase in a chiral honeycomb lattice model}
\author{Genki Yonezawa}
\affiliation{%
Department of physics, Faculty of Science, Kyushu University, Fukuoka, Japan
}%
\author{Jun-ichi Fukuda}
\affiliation{%
Department of physics, Faculty of Science, Kyushu University, Fukuoka, Japan
}%
\affiliation{%
 International Institute for Sustainability with Knotted Chiral Meta Matter (WPI-SKCM$^2$),
Hiroshima University
1-3-1 Kagamiyama, Higashi-Hiroshima, Hiroshima 739-8526, Japan
}%
\author{Toshikaze Kariyado}
\affiliation{%
Reserch Center for Materials Nanoarchitectonics(MANA), National Institute for Materials Science, Tsukuba, Japan
}%

\begin{abstract}
    We investigate topological properties of a chiral honeycomb lattice model
    with next-nearest-neighbor hoppings characterized by the reflection symmetry breaking.
    Topological nontriviality is detected by analyzing effective Dirac Hamiltonian, and confirmed by numerical and analytical study of the emergence of topological edge states at the boundaries between topologically distinct regions. We have also discovered that a novel asymmetric edge current attributable to chirality
    can be excited without any involved phase shifts in input sources to pick up one of the pseudospin components.
\end{abstract}
\maketitle

\section{Introduction}

Band theory, which specifies band energies and wave functions by momentum in Brillouin zone, 
has been traditionally used in solid state physics. Historically, band gaps or effective masses, which are encoded in band energies, play important roles in relation to the semiconductor technology. The study of quantum Hall effect (QHE) \cite{PhysRevLett.45.494, PhysRevLett.49.405}
then brought a new approach to classification of band structures, 
namely the use of topological invariants like Chern numbers that are encoded in the connection of the wave functions in the Brillouin zone. 
Although the QHE requires the time-reversal symmetry breaking, 
it has been recognized that various systems with time-reversal symmetry
can also exhibit topological phases, realized
by spin-orbit coupling \cite{PhysRevLett.95.146802, PhysRevB.76.045302, doi:10.1126/science.1133734}, 
crystalline symmetry \cite{PhysRevLett.106.106802}, and so on. 

As an ideal platform for investigating topological phases,
tight-binding models on honeycomb lattice
have been attracting interest as 
a system exhibiting the Dirac band structure.
One of the most interesting model
was proposed by Haldane \cite{PhysRevLett.61.2015}, 
and a quantum anomalous Hall effect (QAHE) can be realized 
by introducing next-nearest-neighbor (NNN) hopping with complex value.
These complex hoppings can also be induced by considering
the intrinsic spin-orbit coupling (SOC)
in the honeycomb lattice, and spinful electron systems show a quantum spin
Hall effect (QSHE) \cite{RevModPhys.82.3045, RevModPhys.83.1057, PhysRevLett.95.226801}.
After these theoretical predictions of QAHE and QSHE,
many studies have been established to realize topological phases on the honeycomb lattice \cite{Guinea:2010aa,Gomes:2012ab,Hunt:2013aa,Yan:2013aa,PhysRevB.82.161414,PhysRevB.90.075114,Liang_2013,PhysRevB.87.155415}. 
QAHE has also been explored to support 
topological states in metamaterial settings, such as 
Floquet systems \cite{PhysRevB.79.081406, PhysRevB.84.235108, PhysRevLett.111.185307} or 
photonic crystals \cite{PhysRevA.78.033834, PhysRevLett.100.013905}.

As another direction of study, 
a modulated honeycomb lattice model proposed by Wu and Hu \cite{QpSHE}
should be mentioned as a system exhibiting a quantum pseudospin
Hall effect (QpSHE).
They take a hexagonal unit cell
and treat the honeycomb lattice as 
a triangle network of hexagons.
By tuning the ratio of intra-hexagon hopping
to inter-hexagon hopping, a topological phase 
transition can be realized, accompanied by a band inversion
at $\Gamma$ point.
This model has wide application to metamaterials, because
it does not require SOC.
In fact, their setup is also applicable to a photonic crystal made of only a dielectric medium \cite{PhysRevLett.114.223901, doi:10.1126/sciadv.aaw4137}. 

In the studies of honeycomb lattice models, the sublattice symmetry often plays an important role. Only with the nearest-neighbor (NN) hoppings (which is supposed to give a minimal model for graphene), a honeycomb lattice model 
has the sublattice symmetry.
The sublattice symmetry is also preserved in the case of QpSHE in the honeycomb lattice when there is no NNN hoppings. In contrast, the honeycomb models for QAHE and QSHE breaks
the sublattice symmetry due to the complex NNN hoppings \cite{topological-classification, topological-classification-2}.

Although NNN hopping on honeycomb lattice brings exotic phenomena, the previous study of the QpSHE 
did not focus on how the sublattice symmetry breaking affects the topological phase transition.
Moreover, the effect of chirality characterized by the reflection symmetry breaking has not been explored.


The aim of the present study is to investigate how chirality affects the
classification of the band topology and the edge transport
characteristic of the non-trivial band topology.
We focus on the cases in which time reversal symmetry is preserved.
As mentioned above,
a tight-binding model with the honeycomb structure has been regarded as an
ideal platform for topological phases, and here we propose a chiral
honeycomb lattice model by extending the model proposed by Wu and Hu \cite{QpSHE} and
introducing chirality in the NNN hopping. The setup of our model also
breaks the above-mentioned sublattice symmetry, which complements the previous studies of the QpSHE.

The rest of this paper is organized as follows.
In Sec.II, we start with describing the setup of our tight-binding model.
As we will discuss in detail there, our model is applicable not only to
quantum fermionic system but also a classical system characterized by a
dynamical matrix instead of a quantum Hamiltonian.
Then in Sec.III, we conduct topological classification
by using effective Hamiltonian, and calculate energy dispersions.
The validity of analytically shown topological classification is
numerically confirmed by calculating interface states.
In Sec.V, we calculate interface transport to elucidate the effect of
chirality.
We conclude this paper and make discussions in Sec.VI.

\section{Model}
\label{Sec:Model}
In order to consider how chirality affects the behavior of a topologically non-trivial system, 
we specifically discuss a two-dimensional fermionic system or its corresponding classical harmonic oscillator system. 
Generally, band structures is not only for quantum systems but also for classical systems,
such as frequency spectra of dynamical matrices in a spring-mass model \cite{Kariyado:2015aa}.
Therefore, topological nontriviality of band structures could be detected by using both
fermionic system and harmonic oscillator system.

\begin{figure}[htbp]
  \centering
  \includegraphics[width=0.4\textwidth]{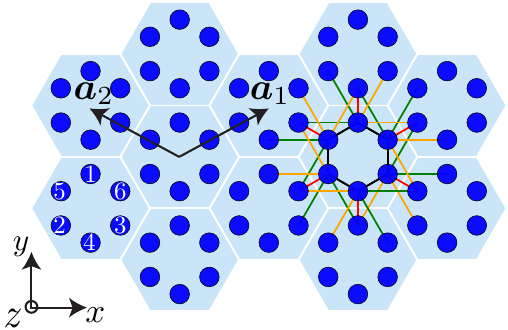}
  \caption{Schematic illustration of the tight-binding model treated in this article.
  The NN hoppings inside unit cells are denoted by $t_0$ (black solid lines),
  and the NN hoppings between unit cells are denoted by $t_1$ (red solid lines).
  The NNN hoppings are also introduced as $t_2$ (green solid lines)
  and $t_3$ (orange solid lines).}
  \label{ft1}
\end{figure}

We first consider a two-dimensional tight-binding model on a honeycomb lattice with NN 
and NNN hoppings. Figure \ref{ft1} presents 
the schematic illustration of the model and the geometry. Due to the modulation in hoppings, the primitive unit cell is a hexagon containing six sites instead of two in the pristine honeycomb lattice model. Then, the unit vectors for the modulated honeycomb lattice are 
$\bm{a}_1 = (3a_0/2, \sqrt{3}a_0/2)^T$ and $\bm{a}_2 = (-3a_0/2, \sqrt{3}a_0/2)^T$, where $a_0$ is a lattice constant for the pristine case.
The Hamiltonian reads
\begin{align}
  H = \sum_{\braket{i, j}}t_{ij}c^\dagger _ic_j + \sum_{\braket{\braket{i', j'}}}t_{i'j'}c^\dagger _{i'}c_{j'}. \label{e1}
\end{align}
Here $c_i$ ($c^\dagger _i$) is the annihilation (creation) operator for electrons.
$\braket{i,j}$ denotes nearest neighbors, and $\braket{\braket{i',j'}}$ denotes next nearest neighbors.
As NN hoppings, we consider both intrahexagon hopping $t_0$ and interhexagon hopping $t_1$,
as depicted in FIG.~\ref{ft1} by black bonds and red bonds respectively.
On the other hand, as NNN hoppings,
we only consider interhexagon hopping.
Furthermore, we classify the interhexagon hoppings into two types, and respectively assign $t_2$ and $t_3$ to introduce chirality on the system. The two types of hoppings are illustrated as green and orange bonds in FIG.~\ref{ft1}.

\begin{figure}[H]
  \centering
  \includegraphics[width=0.48\textwidth]{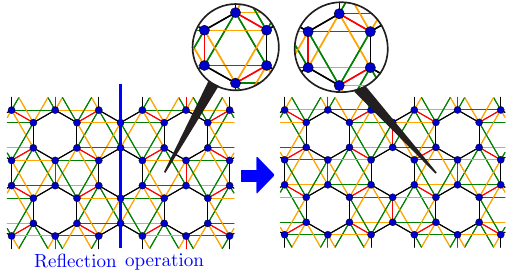}
  \caption{The system after the reflection operation. Due to the existence of NNN hoppings $t_2$
  and $t_3$ the reflected system can not be superposed onto the original one.}
  \label{ft1-1}
\end{figure}

The reflection symmetry breaking can be introduced by setting $t_2$ unequal to $t_3$.
Indeed, as depicted in FIG.~\ref{ft1-1}, the green bonds are the mirror image of the orange bonds, where the reflection plane is on the blue line. This means that the system becomes chiral when $t_2\neq t_3$, 
because the roles of $t_2$ and $t_3$ are swapped before and after the reflection operation.

As mentioned above, this model is an extension of the model proposed by Wu and Hu \cite{QpSHE}, who considered a quantum pseudospin Hall effect on a modulated honeycomb lattice.
Their setup is also applicable to a photonic crystal made of only a dielectric medium \cite{PhysRevLett.114.223901}, which yields a venue for experimental validations \cite{doi:10.1126/sciadv.aaw4137}.

In the following, 
we limit ourselves to the case with the real-valued hopping $t_0, t_1, t_2, t_3$, unlike Haldane model. This makes it straightforward to realize the model in any artificial systems like photonic crystals. Without complex-valued hoppings,  
the time-reversal symmetry is preserved in our setup.
We emphasize that we handle the case of $t_2\neq t_3$, in which the system becomes chiral.

\begin{figure}[H]
  \centering
  \includegraphics[width=0.50\textwidth]{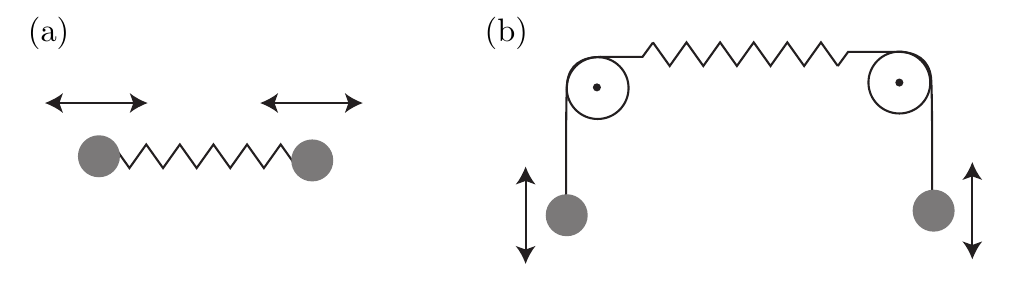}
  \caption{The physical meaning of $s_{ij}$ in a 
    spring-mass model. (a)$s_{ij}=+1$. The masses are connected by a spring directly, then the spring acquires elastic energy for \textit{anti}-phase oscillation. (b)$s_{ij}=-1$. The pulleys change the coordinate axes,
    and $s_{ij}$ is now negative. In this case the spring acquires elastic energy for \textit{in}-phase oscillation.}
  \label{ft4}
\end{figure}

In principle, 
the Hamiltonian Eq.~(\ref{e1}) can be mapped to a dynamical matrix $\Gamma$
of a classical system.
Here we specifically consider a mass-spring system in which potential energy  can be written as 
\begin{align}
  V\{\bm{x}\} = \frac{1}{2}\sum_{i}\sum_{j>i}k_{ij}(x_i-s_{ij}x_j)^2 + \frac{1}{2}\sum_{i}\epsilon_ix_i ^2,
\end{align}
as a function of dynamical variables $\bm{x}$. The number of components in $\bm{x}$ corresponds to the number of degrees of freedom in a given system. The first term represents couplings between different degrees of freedom,  with spring constants $k_{ij}>0$ and $s_{ij}$ being $+1$ or $-1$.
The way to choose $\pm 1$ for $s_{ij}$s in a spring-mass model is explained in FIG.~\ref{ft4}. When a spring stores elastic energy for anti-phase motion of two connected mass points, $s_{ij}=1$ for this spring, while when it stores elastic energy for in-phase motion of two connected mass points, $s_{ij}=-1$ for this spring.
The second term, introduced for later convenience, is a local term that, in a spring-mass model, can be understood as a connection between the mass and
the ground. Here $\epsilon_i$ is positive.
The dynamical matrix of the system is given by
\begin{align}
  \Gamma_{ij} = \frac{\partial^2 V}{\partial x_i\partial x_j} = \left(\epsilon_i + \sum_{l}k_{il}\right)\delta_{ij} - s_{ij}k_{ij}.
\end{align}
By appropriately choosing $\epsilon_i$s, $k_{ij}$s and $s_{ij}$s such that
$\epsilon_i + \sum_{l}k_{il}$ is equal to a constant $\epsilon$ independent of $i$, 
and that $s_{ij}k_{ij}$, which can be positive or negative, is equal to the hopping energies of the quantum counter part,
the dynamical matrix $\Gamma$ can be written as
\begin{align}
  \Gamma_{ij} = \epsilon \delta_{ij} - h_{ij}, \label{eq:map}
\end{align}
where $h_{ij}$ is hopping energies in the Hamiltonian.
By setting $\epsilon$ to be sufficiently large, $\Gamma$ can be positive definite.
Thus, one can construct a classical system where its dynamical matrix $\Gamma$
is equal to $H$ (with a constant shift of $\epsilon$).

After establishing the mapping of Eq.~(\ref{eq:map}), the band structure of 
the Hamiltonian (\ref{e1}) has two interpretations: the energy spectra in the quantum system
and the spectra of the square of the frequency (modified by the constant shift $\epsilon$) 
in the classical system.
For convenience, we investigate the topological properties (Sec.~\ref{Sec:Band} and \ref{Sec:Edge}) by using 
the Hamiltonian $H$, whereas we discuss the interface transport on a classical ribbon structure
with the dynamical matrix $\Gamma$ (Sec.~\ref{Sec:Interfacetransport}), since the coupling with external forces can be understood more intuitively in classical systems.

\section{Band structures and effective Hamiltonian}
\label{Sec:Band}

To elucidate the topological properties of 
our chiral model given by Hamiltonian Eq.~(\ref{e1}), we calculate its energy dispersions.
We first consider the case where the system is periodic with respect to 
$\bm{a}_1$, $\bm{a}_2$. 
By Fourier transforming Eq.~(\ref{e1}), the Hamiltonian as a function of momentum $\bm{k}$ yields
\begin{widetext}
\begin{align}
  &H(\bm{k}) = \begin{pmatrix}
    F & D \\
    D^\dagger & F^T
  \end{pmatrix}, \nonumber \\ 
  &D = \begin{pmatrix}
    t_1e_1 ^*e_2 ^* & t_0 & t_0 \\
    t_0 & t_1e_1 & t_0 \\
    t_0 & t_0 & t_1e_2
  \end{pmatrix},
  F = \begin{pmatrix}
    0 & t_2e_1 ^* + t_3e_1 ^*e_2 ^* & t_2e_1 ^*e_2 ^* + t_3e_2 ^* \\
    t_2e_1 + t_3e_1e_2 & 0 & t_2e_2 ^* + t_3e_1 \\
    t_2e_1e_2 + t_3e_2 & t_2e_2 + t_3e_1 ^* & 0 
  \end{pmatrix}, \label{e3}
\end{align}
\end{widetext}
where $e_l = e^{i\bm{k}\cdot\bm{a}_l}$ $(l=1,2)$.

\begin{figure*}[t]
  \centering
  \includegraphics[width=0.90\textwidth]{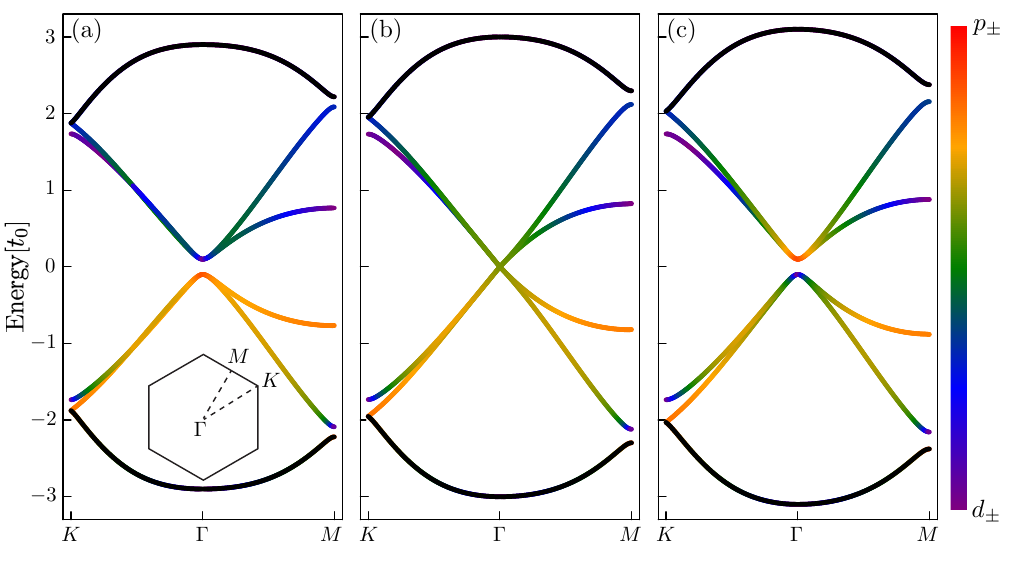}
  \caption{Energy dispersions for the system given by Eq.(\ref{e1}). 
  In all panels, we set $t_2=0.3t_0$ and $t_3=-0.3t_0$:
  (a) $t_1=0.9t_0$, (b) $t_1=t_0$, (c) $t_1=1.1t_0$.
  The color maps are for the values of $\displaystyle |\braket{u^n _{\bm{k}}|d_{+}}|$. $\ket{p_{\pm}}$ and $\ket{d_{\pm}}$ are pseudospin modes, and their definitions are given in Appendix~\ref{Sec:AppendixA}.}
  \label{ft2}
\end{figure*}

Figure~\ref{ft2} shows plots of the energy dispersion for the Hamiltonian Eq.~(\ref{e3})
by setting the hopping energies $t_0$, $t_1$, $t_2$, $t_3$ 
to several typical values.
Importantly, a band inversion occurs by changing the value of
$t_1$ appropriately.
This can be confirmed by plotting the values of $\displaystyle |\braket{u^n _{\bm{k}}|d_{+}}|$ as line colors, 
where $\ket{u^n _{\bm{k}}}$ is the periodic part of Bloch function labeled by index $n$ and Bloch
wave vector $\bm{k}$. We see that the colors of the band edges near $E=0$ at the $\Gamma$-point are exchanged between Figs.~\ref{ft2}(a) and \ref{ft2}(c). 
For the system with $t_1=t_0$ [Fig.~\ref{ft2}(b)], double Dirac cones appear at $E=0$.

In the low energy region around the $\Gamma$ point, the effective Hamiltonian can be derived as
\begin{align}
  \mathcal{H}^{(\mathrm{eff})}(k_x, k_y) \simeq \begin{pmatrix}
    H_+(k_x, k_y) & 0 \\
    0 & H_-(k_x, k_y)
  \end{pmatrix},
\end{align}
where 
\begin{widetext}
\begin{align}
  H_{\pm}(k_x, k_y) = -(t_2+t_3)I + (t_0 - t_1)\sigma_z +  \frac{t_1|\bm{a}_1|}{2}(\pm k_x\sigma_x + k_y\sigma_y). \label{effective_Hamiltonian} 
\end{align}
\end{widetext}
The derivation is given in Appendix~\ref{Sec:AppendixA}.
This Dirac Hamiltonian Eq.~(\ref{effective_Hamiltonian}) clarifies the origin of the band inversion in FIG.~\ref{ft2}, where the band inversion is induced by varing the hopping energy
$t_1$. In the context of the Dirac Hamiltonian, the sign of the Dirac mass $m := t_0-t_1$ can be flipped by changing $t_1$, which explains the origin of the gap and infers that the two states in Figs.~\ref{ft2}(a) and \ref{ft2}(c) are topologically distinct with each other.

For simplicity, we focus on the sign of the mass term $m$ to pick up topological characters of the system, namely we say two states with the opposite signs of $m$ topologically distinct. Strictly speaking, a topological index often requires information of global structure of Bloch wave functions in the entire Brillouin zone (as the Chern number), not only information of the band order at a single momentum ($\Gamma$-point in this case). However, a description by a Dirac equation with spatial modulation in its mass term gives a universal understanding of topologically protected edge/interface states. This will be confirmed in the following analysis. 




\section{Topological edge state}
\label{Sec:Edge}
In this section, we analyze localized states at the boundary between two regions with distinct topology.
We begin with the analytical approach using the low-energy effective Dirac theory.
Let us consider a case where the periodic boundary condition is 
imposed only in the $x$ direction.
There is a boundary normal to the $y$ direction where the sign of the mass term switches:
$t_0-t_1=m_0>0$ for $y>0$ and $t_0-t_1=-m_0$ for $y<0$.
In the $x$ direction,
$k_x$ is a good quantum number because of the existence of the periodic
boundary condition. In the $y$ direction, however,
we apply a continuous approximation by replacing $k_y$ with $-i\partial_y$ to take into account the spatial dependence of $m$.
The eigenvalue equation of $H_{+}(k_x, k_y)$ then becomes
\begin{widetext}
\begin{align}
  \begin{pmatrix}
    -(t_2+t_3)+m_0\mathrm{sgn}(y) & v(k_x-\partial_y) \\
    v(k_x+\partial_y) & -(t_2+t_3)-m_0\mathrm{sgn}(y) 
  \end{pmatrix}
  \begin{pmatrix}
    \phi_1 \\
    \phi_2
  \end{pmatrix}
  = E\begin{pmatrix}
    \phi_1 \\
    \phi_2
  \end{pmatrix},
\end{align}
\end{widetext}
where $v:=t_1|\bm{a}_1|/2$.
Rewriting this equation in a new basis $\phi_{\pm}=\phi_1\pm\phi_2$, 
the eigenvalue equation yields
\begin{widetext}
\begin{align}
  \begin{pmatrix}
    -(t_2+t_3)+vk_x & m_0\mathrm{sgn}(y)+v\partial_y \\
    m_0\mathrm{sgn}(y)-v\partial_y & -(t_2+t_3)-vk_x
  \end{pmatrix}
  \begin{pmatrix}
    \phi_+ \\
    \phi_-
  \end{pmatrix}
  = E\begin{pmatrix}
    \phi_+ \\
    \phi_-
  \end{pmatrix}. \label{e14}
\end{align}
\end{widetext}
When $m_0/v>0$, the solution obtained under the conditions
that the wavefunction converges at $y=\pm\infty$ and is continuous at $y=0$
is
\begin{widetext}
\begin{align}
  E_+ = -(t_2+t_3) - vk_x, \begin{pmatrix}
    \phi_+ \\
    \phi_-
  \end{pmatrix}
  \propto \begin{pmatrix}
    0 \\
    \mathrm{exp}(-(m_0/v)|y|)
  \end{pmatrix}. \label{e15}
\end{align}
\end{widetext}
The eigenvalue equation for $H_-$ is obtained by simply replacing $k_x$ of
Eq.(\ref{e14}) by $-k_x$.
The eigenenergy and states become
\begin{widetext}
\begin{align}
  E_- = -(t_2+t_3) + vk_x, \begin{pmatrix}
    \phi_+ \\
    \phi_-
  \end{pmatrix}
  \propto \begin{pmatrix}
    0 \\
    \mathrm{exp}(-(m_0/v)|y|)
  \end{pmatrix}.
\end{align}
\end{widetext}
Thus, by solving the eigenvalue equations
of $\displaystyle \mathcal{H}^{(\mathrm{eff})}$, 
one obtain the solutions 
such that the eigenenergies $E_{\pm}$ intersect linearly at $E_0:=-(t_2+t_3)$.
In addition, the corresponding eigenstates are exponentially localized at the boundary  
$y=0$. These localized states are protected by the difference of topology,
i.e. the difference of the sign of mass term in Eq.~(\ref{e12}), which confirms the usefulness of the mass-term based topological classificaiton.

\begin{figure*}[t]
  \centering
  \includegraphics[width=0.9\textwidth]{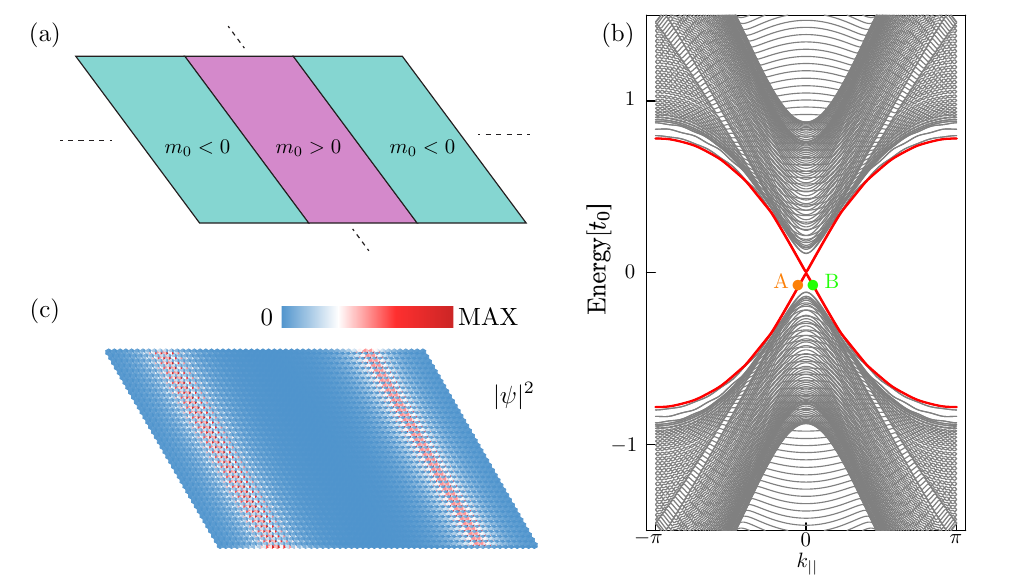}
  \caption{The hopping enegies of chirality are set to $t_2=0.3t_0$ and $t_3=-0.3t_0$.
  (a) Schematic illustration of the system being simulated, with a region of $t_1=1.1t_0$ bounded by two regions of $t_1=0.9t_0$.
  (b) energy dispersions (c) The square of wave functions corresponding to A and B in (b). 
  They have the same value because of time reversal symmetry.}
  \label{ft3}
\end{figure*}

Next, we move on to the numerical approach using the tight-binding model.
In order to discuss boundary states,  
we consider a system where a region of $t_1=1.1t_0$
is sandwiched between two regions of $t_1=0.9t_0$ as shown in FIG.~\ref{ft3}(a). Note that $\bm{a}_1$ direction is horizontal in FIG.~\ref{ft3}(a). Then, periodic boundary conditions are imposed on
$\bm{a}_1$ and $\bm{a}_2$ directions, respectively.
With this construction of the interface, $\bm{a}_2$ remains to be a unit vector, i.e., there is no superstructure in $\bm{a}_2$ direction, and we can calculate energy dispersion as a function of the momentum along the interface $k_\parallel$. The calculated dispersion is in FIG.~\ref{ft3}(b), showing new states in the bulk gap that intersect linearly at $k_\parallel=0$, which is consistent with the analytical discussions above. 
Plotting the square norm of the corresponding wave function (FIG.~\ref{ft3}(c))
also shows that it is localized at the boundaries.

\section{Interface transport}
\label{Sec:Interfacetransport}

So far we have seen that the effect of chirality does not manifest itself in the topological classification or the existence of 
topological edge states, because unequal $t_2$ and $t_3$ just lead to the constant shift of the eigenvalues of the effective Hamiltonian.
In this section, we discuss how chirality affects the interface transport by considering the dynamics of a classical harmonic-oscillator system with boundaries at which energy is injected \cite{PhysRevResearch.3.L032035}. One reason for using a classical system rather than a quantum one is that the meaning of the energy injection can be much more intuitively caught in a classical system as shown below (the energy injection is simply modeled by forced oscillation). Another reason is that the study of modulated honeycomb lattice model has been greatly developed in the context of classical systems like photonic crystals.

To calculate the interface transport,
we denote dynamical variables 
by $\bm{x}=\{x_i\}$. The equation of motion with respect to time $\tau$ is written as
\begin{align}
\frac{d^2 \bm{x}(\tau)}{d\tau^2} = -\Gamma\bm{x}(\tau) + \bm{f}^{c}\cos\Omega \tau. \label{e21}
\end{align}
Here $\Gamma$ is a positive-definite dynamical matrix.
An external force $\displaystyle \bm{f}^{c}\cos\Omega \tau$ is 
applied to the system with a frequency $\Omega$, and the phase of the force is set to be the same at all sites. One can consider a variety of systems including spring-mass systems and LC-circuit systems, and the variables $\bm{x}$ and $\tau$ and the parameters are set dimensionless after rescaling of the length and time according to the specific system to be considered.

To facilitate the physical interpretation, 
let us solve Eq.~(\ref{e21}) by using the normal mode decomposition \cite{PhysRevResearch.3.L032035}.
By introducing an orthogonal matrix $O$ that diagonalizes
$\Gamma$, Eq.~(\ref{e21}) becomes
\begin{align}
  \frac{d^2 \tilde{\bm{x}}(\tau)}{d\tau^2} = -G\tilde{\bm{x}}(\tau) + \tilde{\bm{f}}^{(c)}\cos\Omega \tau,  \label{e22}
\end{align}
where $G = O^T\Gamma O$, $\displaystyle \tilde{\bm{x}}(\tau) = O^T\bm{x}(\tau)$ and
$\tilde{\bm{f}}^{(c)} = O^T \bm{f}^{(c)}$.
The expression of Eq.~(\ref{e22}) in component form becomes
\begin{align}
  \frac{d^2 \tilde{x_l}(\tau)}{d\tau^2} = -\omega_l ^2\tilde{x_l}(\tau) + \tilde{f_l}^{(c)}\cos\Omega \tau, \label{e23}
\end{align}
where we label the normal modes by an integer $l$. 
The solution of Eq.~(\ref{e23}) for the initial condition $\bm{x}(\tau_{\mathrm{init}})=d \bm{x}(\tau_{\mathrm{init}})/d\tau = 0$ is
\begin{align}
  \tilde{x_l}(\tau) = a^{(c)} _l \cos\Omega \tau + b^{(c)} _l \cos\omega_l \tau, \label{e24}
\end{align}
where
\begin{align}
  a^{(c)} _l= \frac{\tilde{f_l}^{(c)}}{\omega_l ^2 - \Omega^2}, b^{(c)} _l = -a^{(c)} _l. \label{e25}
\end{align}
By reconstructing $\bm{x}(\tau)$ from these $\tilde{x}_l(\tau)$,
one can understand the dynamics of the system.
\begin{figure*}[t]
  \centering
  \includegraphics[width=1.0\textwidth]{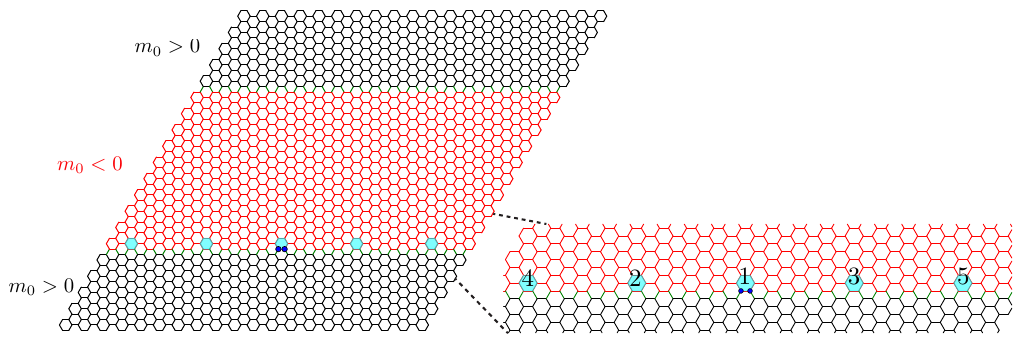}
  \caption{Schematic illustration of the system
  consisting of regions with different topology.
  we set $t_2=-t_3=0.3$ in all region.
  For the mass term part, 
  we set $m_0=-0.2$ for the $m_0<0$ region and $m_0=0.2$ for the $m_0>0$ region.
  $20$ unit cells are lined up in the $\bm{a}_1$ and $\bm{a}_2$ directions with periodic boundary condition.
  We apply the external force $\bm{f}^{(c)}$ at the points of the boundary indicated by blue dots with the 
  amplitude $f_0=1$.}
  \label{ft5}
\end{figure*}

Now, let us return to the interface transport, and consider a specific system illustrated in FIG. \ref{ft5}. 
We consider a ribbon structure of $10$ hexagonal unit cells with $t_1 = 1.2$
cladded from both sides by $5$ hexagonal unit cells with $t_1 = 0.8$.
We also set $t_2=-t_3=0.3$ in all region.
To analyze the dynamics, we construct the dynamical matrix of this ribbon structure
by the procedure Eq.~(\ref{eq:map}), and calculate time evolutions of the intensity at each site $i$
defined as
\begin{align}
  I_i = \frac{1}{2}\Bigl[\Omega^2x_i ^2 + \left(\frac{dx_i}{d\tau}\right)^2\Bigr],\label{eq:intensity}
\end{align}
using Eq.~(\ref{e21}).
Equation~(\ref{eq:intensity}) is designed to eliminate the fast oscillation and focus on effective propagation of energy in a long time scale. 
We also choose $\epsilon = 3.1$ and $\Omega=\sqrt{\epsilon}$.
This choice makes $\Gamma$ positive definite, and $\Omega$ is 
in the bulk gap of the dynamical matrix $\Gamma$.

\begin{figure*}[t]
  \centering
  \includegraphics[width=0.9\textwidth]{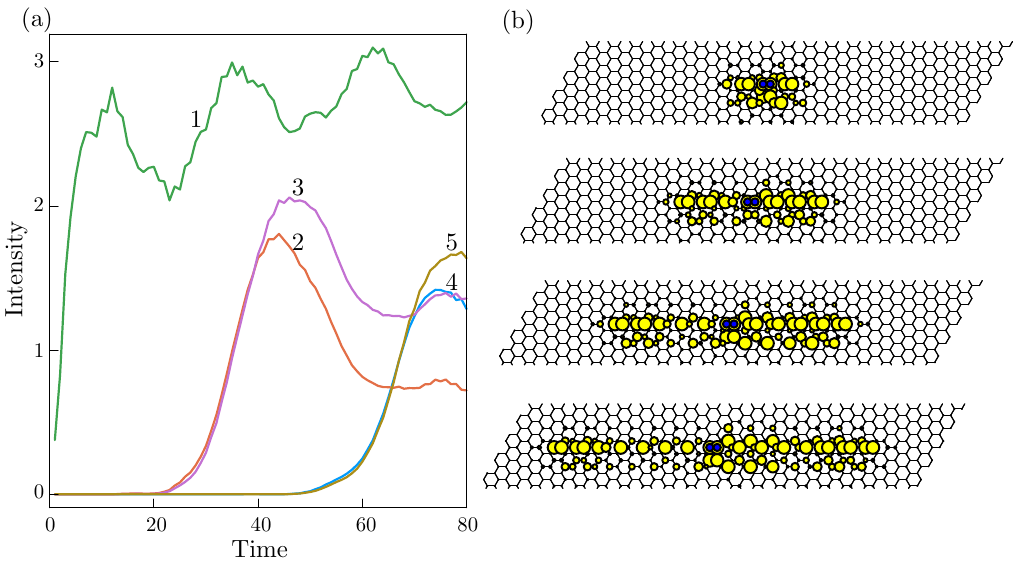}
  \caption{(a) Time evolution of Intensity. The labels 1 to 5 correspond 
  to those in FIG.\ref{ft5}. Intensity in this figure is the sum over the six
  sites in the blue hexagon in FIG.\ref{ft5}.
  (b) Real space plot of the propagation of Intensity $I_i$. 
  We show snapshots at $t=15, 30, 45, 60$.
  An asymmetric propagation is achieved due to chirality.}
  \label{ft6}
\end{figure*}

The obtained results are shown in FIG.~\ref{ft6}. In FIG.~\ref{ft6}(a), the lines labeled by 1-5 correspond to the intensities in the unit cells 1-5 defined in FIG.~\ref{ft5}. The energy is injected at the unit cell 1, and the unit cells 2 and 3 (and also 4 and 5) are located at equal distances from the unit cell 1. The obtained difference between the intensities at 2 and 3 (4 and 5) reveals an asymmetric energy propagation caused by the effect of chirality.
This novel asymmetry is prohibited in the achiral case $t_2=t_3=0$.
It should be noted that $t_2$ and $t_3$ are parameters
that control the asymmetry. If the values of
$t_2$ and $t_3$ are interchanged, 
an asymmetric energy propagation in the opposite direction is observed.
In addition, as can be seen from Eqs.~(\ref{e24}) and (\ref{e25}),
the contribution of the modes with $\omega_l$ close to the frequency
$\Omega$ dominates the energy propagation.
The chirality effects of $t_2$ and $t_3$ manifest
themselves in $\bm{x} (\tau)$ and 
intensity $I_i$ as a linear combination of these multiple modes.

In the achiral case found in the literature, to select right-moving or left-moving interface states, the phase shifts are required in input terminals \cite{Li:2018aa}. In our case, no phase shift is assumed in the forced oscillation term, which gives another way to control energy propagation.

\section{Conclusion and discussion}
We have proposed a universal tight-binding model for chiral two-dimensional systems.
We have shown that the topological classification can be conducted
by using the Dirac Hamiltonian, and confirmed 
the emergence of topological edge states.
We have also discovered a novel asymmetric edge current induced without any phase tuning at input terminals.
These consideration could be a building blocks
in exploring chiral topological materials.

As noted in Sec.\ref{Sec:Interfacetransport}, experimental realization of the system proposed in the present study 
would be more promising in classical systems because the fabrication of 
the system and the tuning of the interactions will be easier than for 
quantum systems. Photonic crystal of a dielectric material
\cite{PhysRevLett.114.223901, Barik_2016} 
could be one direction, and top-down fabrication 
techniques now enable the realization of topological photonics in the 
visible wavelength regime \cite{Liu:2020aa}. Careful fabrication of chiral structures would 
provide an ideal platform for the experimental verification of the 
asymmetric transport discussed in Sec. 5. Self-organization of chiral 
soft materials could also be used for the preparation of systems that 
allows the investigation of chiral transport phenomena, and their ideal 
conditions could be achieved by the tunability of the structural 
properties of the soft material by external stimuli. Recently 
self-assembly of bowtie-shaped nanoparticles has been shown to exhibit 
tunable chiral photonic properties \cite{Kumar:2023aa}. 
Chiral liquid crystals \cite{ChiralityinLiquidCrystals} can also offer a 
platform for self-organized tunable chiral structures, and a hexagonal 
lattice of skyrmions exhibited by a chiral liquid crystal \cite{liquid-crystal} could host asymmetric transport 
phenomena in the visible wavelength regime. Classical mechanics of 
course provides a clue to the realization of systems for chiral 
topological transport phenomena, such as mass-spring systems discussed 
in Sec.\ref{Sec:Interfacetransport} \cite{Kariyado:2015aa}, and spinning top systems \cite{Nash:2015aa}. 
We hope that the present study 
will stimulate experimental studies towards the realization and 
observation of asymmetric topological transport phenomena in a system 
with time-reversal symmetry.
\section*{Acknowledgements}
G. Y. is supported by the Kyushu University Leading Human Resources Development Fellowship Program.
This study is also supported by JSPS KAKENHI No. JP21H01049 (J. F.) and No. JP20K03844 (T. K.).

\appendix
\section{Derivation of the effective Hamiltonian}
\label{Sec:AppendixA}
To derive the effective Hamiltonian,
we focus on the eigenstates at the $\Gamma$ point, i.e. $\bm{k}_\Gamma = \bm{0}$. The following derivation is in parallel with the supplementary materials of the previous study \cite{PhysRevResearch.1.032027}, but any terms with $t_2$ and $t_3$ are new. 
The eigenstates of $H(\bm{k}=\bm{k}_\Gamma)$ are given by
\begin{align}
  \ket{f_{y(3x^2-y^2)}} &= (-1, -1, -1, 1, 1, 1)^T/\sqrt{6}, \nonumber \\
  \ket{p_x} &= (0, -1, 1, 0, 1, -1)^T/2, \nonumber \\
  \ket{p_y} &= (2, -1, -1, -2, 1, 1)^T/2\sqrt{3}, \nonumber \\
  \ket{d_{x^2-y^2}} &= (-2, 1, 1, -2, 1, 1)^T/2\sqrt{3}, \nonumber \\
  \ket{d_{xy}} &= (0, 1, -1, 0, 1, -1)^T/2, \nonumber \\
  \ket{s} &= (1, 1, 1, 1, 1, 1)^T/\sqrt{6}. 
\end{align} \label{e4}
We use the conventional notation of $s, p, d$ and $f$ atomic orbitals.
The corresponding eigenenergies are 
$E_{f_{y(3x^2-y^2)}} =-2t_0-t_1+2t_2+2t_3$, $E_{p_x, p_y} = t_0-t_1-t_2-t_3$, 
$E_{d_{x^2-y^2}, d_{xy}} = t_1-t_0-t_2-t_3$ and $E_s = 2t_0+t_1+2t_2+2t_3$, respectively. 
In the following, 
we consider the case $E_{f_{y(3x^2-y^2)}} < E_{p_x, p_y, d_{x^2-y^2}, d_{xy}} < E_s$.
Based on these eigenstates, one can construct
a low-energy effective Hamiltonian around the
$\Gamma$ point.
Since we focus on the neighborhood of the
$\Gamma$ point where the bands are dominated by $p$ and $d$ states,
it is sufficient to use $\{\ket{p_x}, \ket{p_y}, \ket{d_{x^2-y^2}}, \ket{d_{xy}}\}$ as the basis
in calculating the effective Hamiltonian.
By using these four eigenstates, 
we define the following pseudospin modes:
\begin{align}
  \ket{p_{\pm}} &= \frac{1}{\sqrt{2}}(\ket{p_x}\pm i\ket{p_y}), \\
  \ket{d_{\pm}} &= \frac{1}{\sqrt{2}}(\ket{d_{x^2-y^2}}\pm i\ket{d_{xy}}).
\end{align}


In order to consider the effective Hamiltonian, 
it is convenient to introduce the following basis:
\begin{align}
  \ket{u_{\pm}} &= \frac{1}{\sqrt{2}}(\mp i\ket{p_{\pm}}-\ket{d_{\mp}}), \\
  \ket{l_{\pm}} &= \frac{1}{\sqrt{2}}(i\ket{p_{\pm}}\mp\ket{d_{\mp}}).
\end{align}
Here, the explicit expression of $\{\ket{u_-}, \ket{u_+}, \ket{l_-}, \ket{l_+}\}$ is
\begin{align}
  \ket{u_{\pm}} = \begin{pmatrix}
    \ket{\pm} \\
    0
  \end{pmatrix}, 
  \ket{l_{\pm}} = \begin{pmatrix}
    0 \\
    \pm\ket{\pm}
  \end{pmatrix}, 
\end{align}
where
\begin{align}
    \ket{\pm} = \begin{pmatrix}
    1 \\
    \omega_{\pm} \\
    \omega_{\mp} 
  \end{pmatrix},
  \omega_{\pm} = -\frac{1}{2} \pm \frac{\sqrt{3}}{2}i.
\end{align}
Now, let us first calculate the low-energy effective Hamiltonian by using
$\{\ket{u_-}, \ket{u_+}, \ket{l_-}, \ket{l_+}\}$. 
Indeed, if we expand the Hamiltonian in the basis, the effective Hamiltonian becomes
\begin{widetext}
\begin{align}
  \mathcal{H}^{(\mathrm{eff})}(k_x, k_y) = \begin{pmatrix}
    \bra{-}F\ket{-} & \bra{-}F\ket{+} & -\bra{-}D\ket{-} & \bra{-}D\ket{+} \\
    \bra{+}F\ket{-} & \bra{+}F\ket{+} & -\bra{+}D\ket{-} & \bra{+}D\ket{+} \\
    -\bra{-}D^\dagger \ket{-} & -\bra{-}D^\dagger \ket{+} & \bra{-}F^T\ket{-} & \bra{-}F^T\ket{+} \\
    \bra{+}D^\dagger \ket{-} & \bra{+}D^\dagger \ket{+} & \bra{+}F^T\ket{-} & \bra{+}F^T\ket{+}
  \end{pmatrix}.
\end{align}
\end{widetext}
Then, we perform the Taylor expansion up to the first order of the wavevectors.
The effective Hamiltonian is approximated as 
\begin{widetext}
\begin{align}
  \mathcal{H}^{(\mathrm{eff})}(k_x, k_y) \simeq -(t_2+t_3)I\otimes I + (t_0-t_1)\sigma_x\otimes\sigma_z + \frac{t_1|\bm{a}_1|}{2}\sigma_x\otimes(\bm{k}\cdot\bm{\sigma}), \label{e9}
\end{align} 
\end{widetext}
where $I$ is the identity matrix and $\sigma_i (i=x, y, z)$ is
the Pauli matrices. We denote the Kronecker product as $\otimes$.
It is shown that Eq.(\ref{e9}) can be written as the Dirac Hamiltonian
by introducing the following new basis:
\begin{align}
  \ket{1} &= i\ket{p_-} = \frac{\ket{u_-}+\ket{l_-}}{\sqrt{2}}, \\
  \ket{2} &= -\ket{d_-} = \frac{\ket{u_+}+\ket{l_+}}{\sqrt{2}}, \\
  \ket{3} &= -i\ket{p_+} = \frac{\ket{u_+}-\ket{l_+}}{\sqrt{2}}, \\
  \ket{4} &= -\ket{d_+} = \frac{\ket{u_-}-\ket{l_-}}{\sqrt{2}}.
\end{align}
Using these new bases, the effective Hamiltonian is rewritten as 
\begin{align}
  \mathcal{H}^{(\mathrm{eff})}(k_x, k_y) \simeq \begin{pmatrix}
    H_+(k_x, k_y) & 0 \\
    0 & H_-(k_x, k_y)
  \end{pmatrix},
\end{align}
where 
\begin{widetext}
\begin{align}
  H_{\pm}(k_x, k_y) = -(t_2+t_3)I + (t_0 - t_1)\sigma_z +  \frac{t_1|\bm{a}_1|}{2}(\pm k_x\sigma_x + k_y\sigma_y). \label{e12} 
\end{align}
\end{widetext}
Thus, we obtain the Dirac Hamiltonian Eq.(\ref{e12}) with the constant energy shift $ -(t_2+t_3)$.

Let us make a remark about the effect of $t_2$, $t_3$
from the point of topological classification.
When $t_2 = t_3 = 0$, the original Hamiltonian
proposed by Wu and Hu \cite{QpSHE} is retrieved.
The effective Hamiltonian Eq.(\ref{e12}) 
becomes
\begin{align}
  H_{\pm}(k_x, k_y) = (t_0 - t_1)\sigma_z +  \frac{t_1|\bm{a}_1|}{2}(\pm k_x\sigma_x + k_y\sigma_y). 
\end{align}
Therefore, topological classification by using mass term $m:=t_0 - t_1$ survive in this case.
Moreover, the matrix $F$ in Eq.(\ref{e3}) becomes the zero matrix,
and the Hamiltonian has
the sublattice symmetry
\begin{align}
    \gamma H(\bm{k})\gamma^\dagger = - H(\bm{k}), \gamma^2=1
\end{align}
with
\begin{align}
    \gamma = \begin{pmatrix}
        1 & 0 & 0 & 0 & 0 & 0 \\
        0 & 1 & 0 & 0 & 0 & 0 \\
        0 & 0 & 1 & 0 & 0 & 0 \\
        0 & 0 & 0 & -1 & 0 & 0 \\
        0 & 0 & 0 & 0 & -1 & 0 \\
        0 & 0 & 0 & 0 & 0 & -1 \\
    \end{pmatrix}.
\end{align}
In this case, thanks to this sublattice symmetry, one can define mirror winding number \cite{Kariyado:2017th}, 
and topological classification can be conducted mathematically rigorously.
In the case $t_2\neq 0$ and $t_3\neq 0$, however, the sublattice symmetry is broken, 
and one can not choose the strategy to use the mirror winding number in classifing topological phase.

\bibliographystyle{unsrt}
\bibliography{chiralTB}

\end{document}